\documentclass[aps,pra,twocolumn,showpacs,floatfix]{revtex4}
\usepackage{graphicx}
\usepackage{dcolumn}
\usepackage{amsmath}
\begin{document}

\title{\bf Relativistic effects in two valence electron atoms and ions and
search for variation of the fine structure constant}
\author{E. J. Angstmann}
\author{V. A. Dzuba}
\email{V.Dzuba@unsw.edu.au}
\author{V. V. Flambaum}
\email{V.Flambaum@unsw.edu.au}
\affiliation{School of Physics, University of New South Wales, 
Sydney 2052,Australia}

\date{\today}

\begin{abstract}
We perform accurate calculations of the dependence of transition frequencies 
in two valence electron atoms and ions on a variation of the fine structure 
constant, $\alpha = e^{2}/\hbar c$. The relativistic Hartree-Fock method is used with many-body
perturbation theory and configuration interaction methods to calculate 
transition frequencies. The results are to be used in atomic-clock-type
laboratory experiments designed to test whether $\alpha$ varies in time.

\end{abstract}
\pacs{PACS: 31.30.Jv, 06.20.Jr 95.30.Dr}
\maketitle

\section{Introduction}
Theories unifying gravity with other interactions allow for the possible 
variation of physical constants (see, e.g. \cite{theo1,theo2,theo3}).
Recent analysis of quasar absorption spectra suggests that the fine
structure constant $\alpha$ might vary in space-time
\cite{quasar1,quasar2,quasar3}. There is an
intensive search for alternative ways to test whether $\alpha$ 
is varying. One of the very promising methods to study
local present-day variation of fundamental constants in time involves
the use of atomic clocks. In particular, optical atomic
clock transitions are suitable to study the possible variation of the fine 
structure constant.
This is because the ratio of the frequencies of the optical transitions
depend on $\alpha$ alone, while the frequencies of the hyperfine transitions
also depend on the nuclear magnetic moments and the electron-proton mass ratio.

Laboratory measurements involve measuring how the difference between two 
frequencies changes with time. To relate a measurement of the change between 
two frequencies to a change in $\alpha$, the relativistic energy shifts are 
needed. The relativistic energy shift describes how a level moves as 
$\alpha$ varies. Two transition frequencies with very different relativistic
energy shifts are the most desirable candidates for precision experiments as 
they will have the largest relative frequency shift between them.

The best limit on local present day variation of the fine structure constant 
published to date was obtained by comparing cesium and rubidium atomic fountain
clocks \cite{Marison}. Experiments have also been carried out comparing cesium 
and magnesium \cite{Godone} and a H-maser compared with a Hg~II 
clock \cite{Prestage}. There are many proposals for the search of variation of 
$\alpha$ in atomic optical transitions, some of which were analyzed previously 
in \cite{Dzuba1,Dzuba2,Dzuba3}. In the present work we perform relativistic 
many-body calculations to find the relativistic energy shift for many two 
valence electron atoms and ions. Two valence electron atoms and ions were 
chosen since many new optical clocks experiments, some of which are currently 
under construction and some still under consideration, utilize these atoms and 
ions (e.g.. Al~II \cite{Wineland}, Ca~I \cite{Udem}, Sr~I 
\cite{Courtillot,Katori,Takamoto}, In~II 
\cite{Becker,Nagourney,Zanthier}, Yb~I, Hg~I \cite{Bize,Porsev}).  

\section{Theory}

In the present work we perform calculations for closed shell atoms and ions 
which can also be considered as atoms/ions with two valence electrons above 
closed shells. We start our calculations from the relativistic Hartree-Fock 
(RHF) (also known as Dirac-Hartree-Fock) method in the $V^{N}$ approximation.
This means that RHF calculations are done for the ground state of the
corresponding atom/ion with all electrons included in the self-consistent
field. The use of the $V^{N}$ RHF approximation ensures good convergence
of the consequent configuration interaction (CI) calculations for 
the ground state.
Good accuracy for excited states is achieved by using a large set of 
single-electron states. Note that there is an alternative approach
which uses the $V^{N-2}$ starting approximation (with two valence electrons
removed from the RHF calculations). This approach has some advantages,
it is simpler, and ground and excited states are treated equally. However,
the convergence with respect to the size of the basis is not as good and the 
final results are better in the $V^N$ approximation. We use the $V^{N-2}$
approximation as a test of the accuracy of calculations of the relativistic
energy shifts, while presenting all results in the $V^N$ approximation.

We use a form of the single-electron wave function that explicitly includes a 
dependence on $\alpha$:
\begin{equation}
  \psi(\textbf{r})_{njlm}=\frac{1}{r}\Big(\begin{array}{c}f(r)_{n}
    \Omega(\textbf{r}/r)_{jlm}\\i\alpha g(r)_{n}\tilde{\Omega}
    (\textbf{r}/r)_{jlm}\end{array}\Big).
\end{equation}
This leads to the following form of the RHF equations (in atomic units):
\begin{eqnarray}\label{eq:HF}
  f^{'}_{n}(r)+\frac{\kappa_{n}}{r}f_{n}(r)-[2+\alpha^{2}(\epsilon_{n}-
    \hat{V}_{HF})]g_{n}(r)=0,  \nonumber\\
  g^{'}_{n}(r)+\frac{\kappa_{n}}{r}g_{n}(r)+(\epsilon_{n}-\hat{V}_{HF})
    f_{n}(r)=0,
\end{eqnarray}
where $\kappa=(-1)^{l+j+1/2}(j+1/2)$, $n$ is the principle quantum number 
and $\hat{V}_{HF}$ is the Hartree-Fock potential. The non-relativistic limit 
corresponds to setting $\alpha = 0$.

We then use the combination of the configuration interaction (CI) method with 
the many-body perturbation theory (MBPT)\cite{CIMBPT,DJ98}. 
Interactions between valence electrons are treated using the CI method while 
correlations between the valence electrons and the core electrons are  
included by means of the MBPT. We can write the effective CI Hamiltonian 
for two valence electrons as:
\begin{equation}\label{eq:HCI}
  \hat{H}^{CI} =\hat h_1+\hat h_2+ \hat h_{12}
\end{equation}
here $\hat{h_{i}}$ ($i=1$ or $2$) is an effective single-electron Hamiltonian given by 
\begin{equation}\label{eq:hi}
  \hat{h_{i}}= c \mbox{\boldmath{$\alpha$}}\times {\mathbf{p}} + 
  (\beta-1)mc^{2}-\frac{Ze^{2}}{r_{i}}+\hat V_{core}+\hat \Sigma_{1},
\end{equation}
$\hat V_{core}$ is the Hartree-Fock potential created by the core electrons, 
it differs from $\hat V_{HF}$ in Eq. (\ref{eq:HF}) by the contribution of the 
valence electrons.  $\hat \Sigma_{1}$ is the one-electron operator that  
describes the correlation interaction between a valence electron and the core. 
The third term in Eq. (\ref{eq:HCI}) describes the interaction of the valence 
electrons with each other and can be written as
\begin{equation}\label{eq:hij}
\hat h_{12}=\frac{e^{2}}{r_{12}}+\hat \Sigma_{2}
\end{equation}
where $\hat{\Sigma_{2}}$ is a two-particle operator that describes the effects of 
screening of the Coulomb interaction between the valence electrons by the core 
electrons. The operators $\hat{\Sigma_{1}}$ and $\hat{\Sigma_{2}}$ 
are calculated using the second order of MBPT.

We use the same set of single-electron basis states to construct two-electron
wave functions for the CI calculations and to calculate $\hat \Sigma$. 
The set is based on the B-spline technique developed by Johnson {\em et al} 
\cite{BS1,BS2,BS3}. We use 40 B-splines in a cavity of radius $R=40a_B$ ($a_B$
is Bohr radius). The single-electron basis functions are linear combinations 
of 40 B-splines and are also eigenstates of the Hartree-Fock Hamiltonian
(in the $V^N$ potential). Therefore, we have 40 basis functions in each
partial wave including the B-spline approximations to the atomic core states.
We use a different number of basis states for the CI wave functions and for
the calculations of $\hat \Sigma$. Saturation comes much faster for the 
CI calculations. In these calculations we use 14 states above the core in 
each partial wave up to $l_{max}=3$. Inclusion of states of higher principal 
quantum number or angular momentum does not change the result. 
To calculate $\hat \Sigma$ we use 30 out of 40 states in each partial wave 
up to $l_{max}=4$. 

The results for the energies are presented in Table \ref{tab:levels}.
We present the energies of the $nsnp$ configuration of two electron atoms/ions
with respect to their ground state $^1S_0 \ ns^2$. The states considered
for atomic clock experiments are $^3P_0$ and  $^3P_1$. However,
we present the result for other states as well for completeness, 
these also make it easier to analyze the accuracy of the calculations.
Also, transitions associated with some of these states are observed in quasar
absorption spectra (e.g., the $^1S_0 - ^1P_1$ transition in Ca).

\begin{table}[bt]
\caption{Energies of the $nsnp$ configuration of two electron atoms calculated
using $H^{CI}$, $H^{CI}+\hat{\Sigma_{1}}$ and 
$H^{CI}+\hat{\Sigma_{1}}+\hat{\Sigma_{2}}$; comparison with experiment 
(cm$^{-1}$)}
\begin{ruledtabular}
\begin{tabular}{c c c c c c}
\label{tab:levels}
Atom/ & State & Experiment &\multicolumn{3}{c}{Theory} \\
 ion & & \cite{Moore} & $\hat H^{CI}$ & $\hat H^{CI}+\hat{\Sigma}_1$ & 
 $\hat H^{CI}+\hat{\Sigma}_{1,2}$ \\
\hline
AlII  & $^{3}P_{0}$ & 37393  &  36403  & 36987  & 37328 \\
      & $^{3}P_{1}$ & 37454  &  36466  & 37053  & 37393 \\
      & $^{3}P_{2}$ & 37578  &  36592  & 37185  & 37524 \\
      & $^{1}P_{1}$ & 59852  &  59794  & 60647  & 60090 \\

CaI   & $^{3}P_{0}$ & 15158  &  13701  & 14823  & 15011 \\ 
      & $^{3}P_{1}$ & 15210  &  13750  & 14881  & 15066 \\
      & $^{3}P_{2}$ & 15316  &  13851  & 14997  & 15179 \\
      & $^{1}P_{1}$ & 23652  &  23212  & 24968  & 24378 \\

SrI   & $^{3}P_{0}$ & 14318  &  12489  & 13897  & 14169 \\ 
      & $^{3}P_{1}$ & 14504  &  12661  & 14107  & 14367 \\
      & $^{3}P_{2}$ & 14899  &  13021  & 14545  & 14786 \\
      & $^{1}P_{1}$ & 21698  &  20833  & 23012  & 22305 \\

InII  & $^{3}P_{0}$ & 42276  &  37825  & 39238  & 42304 \\ 
      & $^{3}P_{1}$ & 43349  &  38867  & 40394  & 43383 \\
      & $^{3}P_{2}$ & 45827  &  41168  & 42974  & 45904 \\
      & $^{1}P_{1}$ & 63034  &  62181  & 64930  & 62325 \\

YbI   & $^{3}P_{0}$ & 17288  &  14377  & 16352  & 16950  \\ 
      & $^{3}P_{1}$ & 17992  &  15039  & 17189  & 17705  \\
      & $^{3}P_{2}$ & 19710  &  16550  & 19137  & 19553  \\
      & $^{1}P_{1}$ & 25068  &  24231  & 27413  & 26654  \\

HgI   & $^{3}P_{0}$ & 37645  &  31864  & 32692  & 37420 \\  
      & $^{3}P_{1}$ & 39412  &  33751  & 34778  & 39299 \\
      & $^{3}P_{2}$ & 44043  &  38155  & 39781  & 44158  \\
      & $^{1}P_{1}$ & 54069  &  50247  & 52994  & 56219 \\

TlII  & $^{3}P_{0}$ & 49451  &  43831  & 43911  & 49865 \\  
      & $^{3}P_{1}$ & 52393  &  47091  & 47350  & 52687 \\
      & $^{3}P_{2}$ & 61725  &  55988  & 56891  & 62263 \\
      & $^{1}P_{1}$ & 75660  &  74291  & 76049  & 74717 \\
\end{tabular}
\end{ruledtabular}
\end{table}

To demonstrate the importance of the core-valence correlations we
include results of pure CI calculations (with no $\hat \Sigma$) as well
as the results in which only $\hat \Sigma_1$ is included but $\hat \Sigma_2$
is not. One can see that the accuracy of pure CI calculations is about
10\% while inclusion of core-valence correlations improves it significantly
to the level of about 1\%. The deviation from experiment of the final 
theoretical energies for the triplet states of all atoms except Yb is not more 
than 1\%. For Yb it is 2\%. The accuracy of the singlet states is about 1\%
for the ions, 3-4\% for CaI, SrI and HgI and 6\% for YbI. The accuracy of the 
fine structure intervals ranges from 2 to 7\%. The accuracy of
calculations for Yb is not as good as for other atoms because the two electron
approximation is a poor approximation for this atom. Electrons from the
$4f$ subshell, which are kept frozen in present calculations, are relatively
easy to excite and corresponding configurations give substantial
contribution to the energy. Note that we do include these excitations
perturbatively, into the $\hat \Sigma$ operator. However, due to their
large contribution, second-order treatment of the excitations from the $4f$
subshell is not very accurate. On the other hand, the CI+MBPT
results for Yb are still much better than pure CI values.

Note also that the CI+MBPT results presented in Table \ref{tab:levels}
are in good agreement with similar calculations in Refs.
\cite{Kozlov1,Kozlov2}.

\section{Results and Discussion}

\begin{table}[tb]
\caption{Calculated $q$ coefficients, for transitions from the ground state,  using $H^{CI}$, $H^{CI}+\hat{\Sigma_{1}}$
 and $H^{CI}+\hat{\Sigma_{1}}+\hat{\Sigma_{2}}$}
\begin{ruledtabular}
\begin{tabular}{c c c c c c}\label{tab:q}
Atom/ion &  State &
  \multicolumn{1}{c}{$\hat H^{CI}$}  & 
  \multicolumn{1}{c}{$\hat H^{CI}+\hat \Sigma_{1}$} &
  \multicolumn{1}{c}{$H^{CI}+\hat\Sigma_{1,2}$} & Other \\
\hline
AlII  & $^{3}P_{0}$ &   138 &   142 &   146 & \\
      & $^{3}P_{1}$ &   200 &   207 &   211 & \\
      & $^{3}P_{2}$ &   325 &   340 &   343 & \\
      & $^{1}P_{1}$ &   266 &   276 &   278 & \\

CaI    & $^{3}P_{0}$ &   108 &   115 &   125 & \\
      & $^{3}P_{1}$ &   158 &   173 &   180 & 230 \cite{Dzuba1} \\
      & $^{3}P_{2}$ &   260 &   291 &   294 & \\
      & $^{1}P_{1}$ &   228 &   238 &   250 & 300 \cite{Dzuba1} \\

SrI    & $^{3}P_{0}$ &   384 &   396 &   443 &      \\
      & $^{3}P_{1}$ &   560 &   609 &   642 &  667 \cite{Dzuba4} \\
      & $^{3}P_{2}$ &   939 &  1072 &  1084 &      \\
      & $^{1}P_{1}$ &   834 &   865 &   924 & 1058 \cite{Dzuba4} \\

InII  & $^{3}P_{0}$ &  3230 &  2932 &  3787 & 4414 \cite{Dzuba3} \\
      & $^{3}P_{1}$ &  4325 &  4125 &  4860 & 5323 \cite{Dzuba3} \\
      & $^{3}P_{2}$ &  6976 &  7066 &  7767 & 7801 \cite{Dzuba3} \\
      & $^{1}P_{1}$ &  6147 &  6103 &  6467 &      \\

YbI   & $^{3}P_{0}$ &  2339 &  2299 & 2714 & \\ 
      & $^{3}P_{1}$ &  3076 &  3238 & 3527 & \\
      & $^{3}P_{2}$ &  4935 &  5707 & 5883 & \\
      & $^{1}P_{1}$ &  4176 &  4674 & 4951 & \\

HgI    & $^{3}P_{0}$ & 13231 &  9513 & 15299 & \\
      & $^{3}P_{1}$ & 15922 & 12167 & 17584 & \\
      & $^{3}P_{2}$ & 22994 & 19515 & 24908 & \\
      & $^{1}P_{1}$ & 20536 & 16622 & 22789 & \\

TlII  & $^{3}P_{0}$ & 14535 & 11101 & 16267 & 19745 \cite{Dzuba3} \\
      & $^{3}P_{1}$ & 18476 & 14955 & 18845 & 23213 \cite{Dzuba3} \\
      & $^{3}P_{2}$ & 32287 & 28903 & 33268 & 31645 \cite{Dzuba3} \\
      & $^{1}P_{1}$ & 28681 & 25160 & 29418 &       \\

\end{tabular}
\end{ruledtabular}
\end{table}

\begin{table}[tb]
\caption{Experimental energies and calculated $q$-coefficients (cm$^{-1}$) 
for transitions from the ground state $ns^{2}$ to the $nsnp$ configurations of two-electron atoms/ions}
\begin{ruledtabular}
\begin{tabular}{c c c c l r  }\label{tab:final}
Atom/Ion & Z & \multicolumn{2}{c}{State} & Energy\cite{Moore} & 
\multicolumn{1}{c}{q} \\
\hline
AlII & 13 & $3s3p$ & $^{3}P_{0}$ & 37393.03 & 146 \\
     &    & $3s3p$ & $^{3}P_{1}$ & 37453.91 & 211 \\
     &    & $3s3p$ & $^{3}P_{2}$ & 37577.79 & 343 \\
     &    & $3s3p$ & $^{1}P_{1}$ & 59852.02 & 278 \\

CaI  & 20 & $4s4p$ & $^{3}P_{0}$ & 15157.90 & 125 \\
     &    & $4s4p$ & $^{3}P_{1}$ & 15210.06 & 180 \\
     &    & $4s4p$ & $^{3}P_{2}$ & 15315.94 & 294 \\
     &    & $4s4p$ & $^{1}P_{1}$ & 23652.30 & 250 \\

SrI  & 38 & $5s5p$ & $^{3}P_{0}$ & 14317.52 &  443 \\
     &    & $5s5p$ & $^{3}P_{1}$ & 14504.35 &  642 \\
     &    & $5s5p$ & $^{3}P_{2}$ & 14898.56 & 1084 \\
     &    & $5s5p$ & $^{1}P_{1}$ & 21698.48 &  924 \\

InII & 49 & $5s5p$ & $^{3}P_{0}$ & 42275    & 3787 \\
     &    & $5s5p$ & $^{3}P_{1}$ & 43349    & 4860 \\
     &    & $5s5p$ & $^{3}P_{2}$ & 45827    & 7767 \\
     &    & $5s5p$ & $^{1}P_{1}$ & 63033.81 & 6467 \\

YbI  & 70 & $6s6p$ & $^{3}P_{0}$ & 17288.44 & 2714 \\
     &    & $6s6p$ & $^{3}P_{1}$ & 17992.01 & 3527 \\
     &    & $6s6p$ & $^{3}P_{2}$ & 19710.39 & 5883 \\
     &    & $6s6p$ & $^{1}P_{1}$ & 25068.22 & 4951 \\

HgI  & 80 & $6s6p$ & $^{3}P_{0}$ & 37645.08 & 15299 \\
     &    & $6s6p$ & $^{3}P_{1}$ & 39412.30 & 17584 \\
     &    & $6s6p$ & $^{3}P_{2}$ & 44042.98 & 24908 \\
     &    & $6s6p$ & $^{1}P_{1}$ & 54068.78 & 22789 \\

TlII & 81 & $6s6p$ & $^{3}P_{0}$ & 49451 & 16267  \\
     &    & $6s6p$ & $^{3}P_{1}$ & 53393 & 18845  \\
     &    & $6s6p$ & $^{3}P_{2}$ & 61725 & 33268  \\
     &    & $6s6p$ & $^{1}P_{1}$ & 75600 & 29418   \\

\end{tabular}
\end{ruledtabular}
\end{table}

In the vicinity of the $\alpha_{0}$, the present day value of $\alpha$, the 
frequency of a transition, $\omega$, can be written as:
\begin{equation}\label{eq:omega}
  \omega = \omega_{0}+qx,
\end{equation}
where $x=(\frac{\alpha}{\alpha_{0}})^{2}-1$, $\omega_{0}$ is the present day 
experimental value of the frequency and the $q$ coefficient is the relativistic
energy shift that determines the frequency dependence on $\alpha$. It is clear 
from the above expression that $q$ coefficients can be described by 
\begin{equation*}
q=\frac{d\omega}{dx}|_{x=0}. 
\end{equation*}
Thus, in order to calculate $q$ coefficients the atomic energy levels of the 
atoms and ions of interest at different values of $x$ need to be calculated. 
The relativistic energy shift $q$ is then calculated using the formulae
\begin{equation}
  q=\frac{\omega(\Delta x) - \omega(-\Delta x)}{2\Delta x}
\end{equation}
and
\begin{equation}
  q=\frac{16(\omega(\Delta x) - \omega(-\Delta x)) -
  2(\omega(2\Delta x) - \omega(-2\Delta x))}{24\Delta x}.
\end{equation}
The second formula is needed to check for non-linear contributions to
$d\omega/dx$. We use $\Delta x = 0.1$ and $\Delta x = 0.125$. The results are
presented in Table \ref{tab:q}.

As for the energies, we use three different approximations to calculate
relativistic energy shifts: (1) pure CI approximation for two valence
electrons, (2) CI with $\hat \Sigma_1$ and (3) CI+MBPT approximation with both
$\hat \Sigma_1$ and $\hat \Sigma_2$ included.
Inclusion of core-valence correlations lead to increased values of the 
$q$-coefficients. This is because the correlation interaction of a valence
electron with the core introduces an additional attraction which increase the
density of the valence electron in the vicinity of the nucleus and thus 
emphasize the importance of the relativistic effects.

Note that $\hat \Sigma_1$ and $\hat \Sigma_2$ are of the same order
and need to be included simultaneously to obtain reliable results.
$\hat{\Sigma_1}$ is much easier to calculate and inclusion of $\hat \Sigma_1$ 
alone often leads to significant improvements of the results 
for the energies (see Table \ref{tab:levels}). 
However, the results for the $q$-coefficients
show that neglecting $\hat \Sigma_2$ may lead to significant loss in
accuracy. Indeed, the results for $q$'s with $\hat \Sigma_1$ alone
are often smaller than those obtained in pure CI and CI+MBPT approximations
and differ from final values by up to 50\%. 
Since neglecting $\hat \Sigma_2$
cannot be justified, we present results without $\hat \Sigma_2$ for 
illustration purpose only. 

The accuracy of the calculation of the 
$q$-coefficients can be estimated by comparing the CI and CI+MBPT results
 calculated in the $V^{N}$ and $V^{N-2}$ approximations and also by 
comparing the final results for the energies 
(including fine structure
intervals) with experimental values.
As one can see from Table \ref{tab:q} inclusion of the core-valence
correlations can change the values of the $q$-coefficients by more than 15\%.
However, the accuracy of the energies improves significantly when
core-valence correlations are included. It is natural to expect that
the final accuracy for the $q$-coefficients is also higher when 
core-valence correlations are included.
Comparison with our previous results also shows some deviation 
on approximately the same level (the largest relative discrepancy is
for Ca where relativistic effects are small and high accuracy is not needed).
Most of this discrepancy can be attributed to the inaccuracy of our
old, less complete calculations. Comparison 
between the energies calculated in the $V^{N}$ and $V^{N-2}$ approximations and the experimental values suggest that 10\% is a reasonable estimate of the accuracy of the present calculations of the relativistic energy shifts for Al II, Ca I and Sr I, 15\% for In II, 25\% for Yb I and 20\% for Hg I and Tl II.

In Table \ref{tab:final} we present final values of the relativistic energy 
shifts together with the experimental energies.

\section*{Acknowledgments}
This work is supported by the Australian Research council.

\bibliography{qc}

\end{document}